\documentclass[showpacs,aps,graphicx,twocolumn]{revtex4}%and
\usepackage{graphicx}

\begin{document}

\title{Deterministic  polarization entanglement purification using time-bin entanglement}

\author{Yu-Bo Sheng$^{1}$\footnote{Email address:
shengyb@njupt.edu.cn} , Lan Zhou$^{1,2}$}
\address{$^1$ Key Lab of Broadband Wireless Communication and Sensor Network
 Technology,Nanjing University of Posts and Telecommunications, Ministry of
 Education, Nanjing, 210003,
 China\\
$^2$College of Mathematics \& Physics, Nanjing University of Posts and Telecommunications, Nanjing,
210003, China\\}
\date{\today }

\begin{abstract}
We present a deterministic  entanglement purification protocol (EPP) working for currently available experiment technique. In
this protocol, we resort to  the robust  time-bin entanglement to purify the
polarization entanglement determinately, which is quite different
from the previous EPPs. After purification, one can
obtain a completely pure maximally entangled pair with the success probability of 100\%, in principle.
As the maximal polarization entanglement is of vice importance in long-distance quantum communication,
this protocol may have a wide application.
\end{abstract}
 \maketitle

\section{Introduction}
The distribution of entanglement states between long distant
locations is essential for quantum
communication \cite{computation1,computation2,teleportation}. For
instance, in order to achieve the faithful
teleportation of unknown quantum
states \cite{teleportation,cteleportation}, or quantum
cryptograph \cite{Ekert91,BBM92,rmp}, people first need to set up a
quantum channel with maximally entangled state. Unfortunately,
the source of entanglement is usually fragile. In a practical
transmission, the interaction between a quantum entanglement system
and the innocent noise of quantum channel always exists, which will
make the maximally entangled state degrade and become a mixed state. A degraded
quantum channel will make the fidelity of the teleportation
degraded, and the key in the quantum cryptograph insecure. Therefore, before performing the
quantum communication, they should recover the degraded entangled states into the maximally
entangled states.

Entanglement purification is the method to obtain the maximally
entangled states from the less-entangled ones. It has been widely used
in quantum repeaters \cite{repeater,DLCZ,zhao,chen,Simon1}. The first
entanglement purification protocol (EPP) was proposed by
Bennett \emph{et al.} in 1996 \cite{Bennett1}.  It was used to purify the Werner state \cite{werner} with
quantum logical gate, i.e. controlled-NOT (CNOT) gate. This
protocol was developed by Deustch \emph{et al.}
with similar quantum logical operations, subsequently \cite{Deutsch}.  The
entanglement purification for multiparticle and high dimension
system have also been proposed \cite{Murao,Horodecki,Yong}. However,
the CNOT gates or similar logical operations are very difficult to
implement in current experiment.

On one hand, in long-distance quantum communication, photon encoded in the polarization degree of freedom
is a best qubit system for its simple manipulation and fast
transmission. In 2001, Pan \emph{et al.} proposed an EPP with linear optics \cite{Pan1}.
They used two polarization beam splitters (PBSs) to substitute the CNOT gates to complete the
parity-check measurement which made the EPP can be easily realized in
experiment. Subsequently, they have demonstrated
entanglement purification for general mixed states of
polarization-entangled photons in 2003 \cite{Pan2}. Later, some other EPPs were proposed,
such as the EPP based on the cross-Kerr nonlinearity \cite{shengpra}, the EPPs for single-photon entanglement \cite{sangouard}, quantum-dot and microcavity system \cite{wangc2,wangc3}, and so on \cite{wangc1,dengfuguo2,peter1,peter2,hyperpurification,hybirdpurification}. These EPPs described above are all based
on the frame of CNOT gates or similar logical operations. They should sacrifice a large number of low quality mixed states to improve
the fidelity of the mixed states.

On the other hand, there is another way to realize polarization entanglement purification. It uses the other
degrees of freedom entanglement to purify the polarization entanglement. In 2002,  Simon and Pan first used the spatial  entanglement to
purify polarization entanglement with linear optics \cite{Simon}. However, their protocol can only correct the bit-flip error. In 2010, Sheng \emph{et al.} proposed a deterministic EPP, resorting
to both spatial and frequency entanglement \cite{shengpra3}. However, their protocol is based on the weak cross-Kerr nonlinearity. It makes this protocol hard to realize.
There are other deterministic EPPs based on the spatial entanglement \cite{shengpra4,lixh,dengfuguo}. Unfortunately, the spatial entanglement has its inherent drawback for the relative phase is sensitive to the length fluctuation \cite{zhao,chen}. Though phase can be well controlled in the experiment, as pointed out by Simon and Pan, it is still unrealistic in a practical long-distance quantum communication \cite{Simon,Pan2}. Actually, using entanglement in other degree of freedom, such as spatial entanglement to purify the polarization entanglement is essentially
the entanglement transformation. The reason is that the entanglement purification is based on the local operation and classical communication (LOCC). It is well known that LOCC cannot create the entanglement. The key idea of such EPP is to transfer the robust entanglement to the fragile entanglement.

Interestingly, qubits encoded in the time-bin degree of freedom are
particularly suitable for long-distance quantum communication
and fundamental experiments. The preparation of
time-bin entangled states \cite{timebin1,timebin2,timebin3,timebin4}, violation of Bell inequalities \cite{timebin5,timebin6}, quantum
key distribution \cite{timebin7}, teleportation \cite{timebin8} were widely discussed. Moreover, Humphreys \emph{et al.}
discussed the linear optical quantum computation in a single spatial mode \cite{timebin9}. One of the good advantages of time-bin entanglement
is that it is a robust form of optical quantum information, especially for transmission in
optical fibers. In 2002, Thew \emph{et al.} have experimental
investigated the robust time-bin qubits for distributed quantum
communication over 11 km\cite{Thew}. In 2004, Marcikic \emph{et al.} also
reported the experimental distribution of time-bin entangled qubits
over 50 km of optical fibers. They demonstrated the violation of the
Clauser-Horne-Shimony-Holt Bell inequality by more than 15 standard
deviations without removing the imperfect detectors \cite{Marcikic}.
Recently, Donohue \emph{et al.} reported their experimental results about the tomographically complete set of time-bin qubit projective measurements
and showed that the fidelity of operations is sufficiently high to violate the Clauser-Horne-Shimony-Holt-Bell
inequality by more than 6 standard deviations \cite{timebin10}.

\begin{center}
\begin{figure}[!h]%[tpb]
\includegraphics[width=9cm,angle=0]{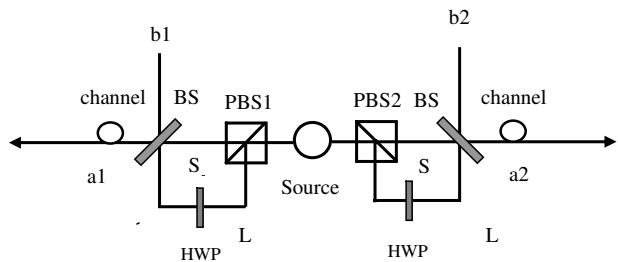}
\caption{Schematic diagram showing the principle of converting the
polarization entanglement to time-bin entanglement. The M-Z
interferometer makes the state be in the different time bins. BS is
the 50:50 beam splitter, and HWP is the half wave plate.}
\end{figure}
\end{center}

In this paper, we present a deterministic EPP resorting to the time-bin entanglement. We still use the concept of the entanglement purification
to describe this protocol. Compared with the conventional EPPs, one can
get a genuine  pure entangled pair without consuming less-entangled
pairs. Moreover, we do not require the initial state to be the hyperentanglement to complete the task.
It will greatly release the complexity of the experiment.
Moreover, the time-bin entanglement is more robust and can be well
manipulated, which makes this EPP more useful in practical application.

\section{Deterministic EPP using time-bin entanglement}
 Now we start to explain this protocol by discussing a simple
example. Interestingly, in this EPP, we do not require the polarization part to be entangled. Therefore, before the parties Alice and Bob share the polarization entangled pair, they first
encode the initial pure polarization entanglement $|\Phi^{+}\rangle$
into the time-bin entanglement. As shown in Fig.1, after PBS1 and PBS2, the
initial state $|\Phi^{+}\rangle$ evolves as
\begin{eqnarray}
|\Phi^{+}\rangle_{AB}&=&\frac{1}{\sqrt{2}}(|H\rangle_{A}|H\rangle_{B}+|V\rangle_{A}|V\rangle_{B})\nonumber\\
&\longrightarrow&\frac{1}{\sqrt{2}}(|H\rangle^{S}_{A}|H\rangle^{S}_{B}+|H\rangle^{L}_{A}|H\rangle^{L}_{B})\nonumber\\
&=&|H\rangle_{A}|H\rangle_{B}\otimes\frac{1}{\sqrt{2}}(|S\rangle_{A}|S\rangle_{B}+|L\rangle_{A}|L\rangle_{B}).\nonumber\\
\end{eqnarray}
The subscripts A and B mean that the photons transmit to Alice and
Bob, respectively. The superscript S and L denote  the different
time bins. The L is the long arm and the S is the short arm. $|H\rangle$ and $|V\rangle$ denote
horizontal and vertical polarizations of the single photon, respectively. After the photons passing through  two 50:50 beam splitters (BSs), the system becomes
\begin{eqnarray}
|\Phi^{+}\rangle_{AB}&=&\frac{1}{2}|H\rangle_{a1}|H\rangle_{a2}\otimes\frac{1}{\sqrt{2}}(|S\rangle_{a1}|S\rangle_{a2}+|L\rangle_{a1}|L\rangle_{a2})\nonumber\\
&+&\frac{1}{2}|H\rangle_{a1}|H\rangle_{b2}\otimes\frac{1}{\sqrt{2}}(|S\rangle_{a1}|S\rangle_{b2}-|L\rangle_{a1}|L\rangle_{b2})\nonumber\\
&+&\frac{1}{2}|H\rangle_{a2}|H\rangle_{b1}\otimes\frac{1}{\sqrt{2}}(|S\rangle_{a2}|S\rangle_{b1}-|L\rangle_{a2}|L\rangle_{b1})\nonumber\\
&+&\frac{1}{2}|H\rangle_{b1}|H\rangle_{b2}\otimes\frac{1}{\sqrt{2}}(|S\rangle_{b1}|S\rangle_{b2}+|L\rangle_{b1}|L\rangle_{b2}).\nonumber\\
\end{eqnarray}
The initial state will be in the different spatial modes with same
probability of $\frac{1}{4}$. Now we only discuss the transmission
in channel a1a2. The same way can be used to the states in the
channel a1b2, a2b1 and b1b2.  The initial state with
$|H\rangle_{a1}|H\rangle_{a2}\otimes\frac{1}{\sqrt{2}}(|S\rangle_{a1}|S\rangle_{a2}+|L\rangle_{a1}|L\rangle_{a2})$
can be rewritten as
\begin{eqnarray}
\rho=\rho_{P}\otimes\rho_{T}.
\end{eqnarray}
Here $\rho_{P}$ is the polarization part with $\rho_{P}=|HH\rangle\langle HH|$, and $\rho_{T}$ time-bin part with
$\rho_{T}=|\Phi\rangle_{T}\langle \Phi|$. Here
$|\Phi\rangle_{T}=\frac{1}{\sqrt{2}}(|S\rangle|S\rangle+|L\rangle|L\rangle)$.
After transmitted in the noisy channel, the
polarization freedom of the photons is incident to be influenced by
the vibration, the thermal fluctuation, or the distort of the fiber.
These effects will lead the polarization entanglement to be degraded,
and make the pure state $\rho_{P}$ become a mixed state $\rho'_{P}$, with
\begin{eqnarray}
\rho'_{P}&=&F|\Phi^{+}\rangle\langle \Phi^{+}|+a|\Phi^{-}\rangle\langle\Phi^{-}|\nonumber\\
&+&b|\Psi^{+}\rangle\langle \Psi^{+}|+c|\Psi^{-}\rangle\langle \Psi^{-}|.
\end{eqnarray}
Here $F+a+b+c=1$, and
\begin{eqnarray}
|\Phi^{-}\rangle&=&\frac{1}{\sqrt{2}}(|H\rangle|H\rangle-|V\rangle|V\rangle),\nonumber\\
|\Psi^{\pm}\rangle&=&\frac{1}{\sqrt{2}}(|H\rangle|V\rangle\pm|V\rangle|H\rangle).
\end{eqnarray}

Fortunately, the time-bin entanglement is more robust than the
polarization entanglement. After the state passing through the noisy
channel, the whole state becomes
\begin{eqnarray}
\rho'=\rho'_{P}\otimes\rho_{T},\label{mix1}
\end{eqnarray}
and arrives at the setup for purification, as shown in Fig. 2.
\begin{widetext}
\begin{center}
\begin{figure}[!h]%[tpb]
\includegraphics[width=16cm,angle=0]{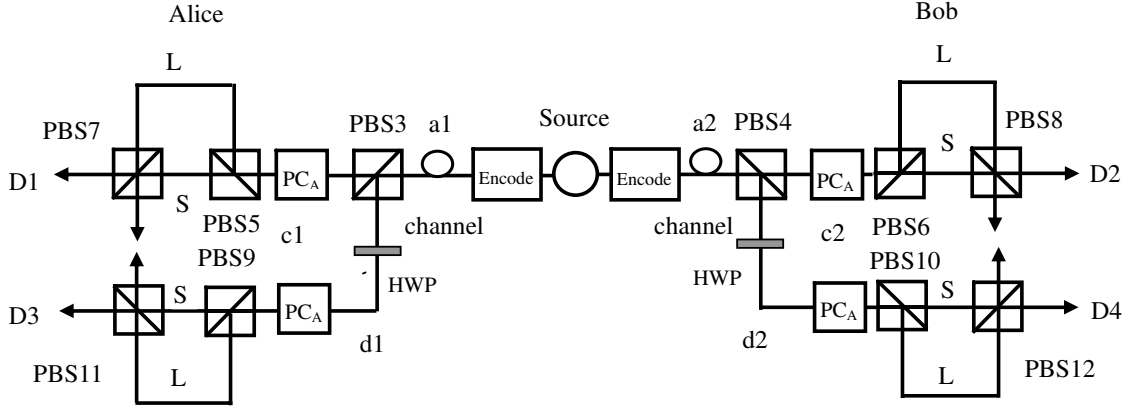}
\caption{Schematic diagram showing the principle of purification.
The PC$_{A}$ is the Pockels cell. The mixed state of  polarization
can be divided into different spatial modes. This setup can purify
the state which comes from a1a2 mode. The states in the other modes
can also be purified by adding the similar setup. If the detectors
D1D2, D1D4, D3D2 or D3D4 click, one can get the maximally entangled
state $|\Phi^{+}\rangle$ determinately. The setup named "Encode" is shown in Fig. 1.}
\end{figure}
\end{center}
\end{widetext}
The state $\rho'$ can be described as: with the probability of $F$, state is in the $|\Phi^{+}\rangle|\Phi\rangle_{T}$.
With the probability of $a$, state is in the $|\Phi^{-}\rangle|\Phi\rangle_{T}$. With the probability of $b$ and $c$, states are
in $|\Psi^{+}\rangle|\Phi\rangle_{T}$ and $|\Psi^{-}\rangle|\Phi\rangle_{T}$, respectively.

We consider the  case with  $|\Phi^{+}\rangle|\Phi\rangle_{T}$. It can be written
as $\frac{1}{2}(|H\rangle_{a1}|H\rangle_{a2}+|V\rangle_{a1}|V\rangle_{a2})(|S\rangle_{a1}|S\rangle_{a2}+|L\rangle_{a1}|L\rangle_{a2})$. We first discuss the item $|H\rangle_{a1}|H\rangle_{a2}\otimes\frac{1}{\sqrt{2}}(|S\rangle_{a1}|S\rangle_{a2}+|L\rangle_{a1}|L\rangle_{a2})$.
After passing through the PBS3 and PBS4,  the whole state evolves as
\begin{eqnarray}
&&|H\rangle_{a1}|H\rangle_{a2}\otimes\frac{1}{\sqrt{2}}(|S\rangle_{a1}|S\rangle_{a2}+|L\rangle_{a1}|L\rangle_{a2})\nonumber\\
&\rightarrow&|H\rangle_{c1}|H\rangle_{c2}\otimes\frac{1}{\sqrt{2}}(|S\rangle_{c1}|S\rangle_{c2}+|L\rangle_{c1}|L\rangle_{c2})\nonumber\\
&\rightarrow&\frac{1}{\sqrt{2}}(|V\rangle_{c1}|V\rangle_{c2}\otimes|S\rangle_{c1}|S\rangle_{c2}
+|H\rangle_{c1}|H\rangle_{c2}\otimes|L\rangle_{c1}|L\rangle_{c2})\nonumber\\
&=&\frac{1}{\sqrt{2}}(|H\rangle^{L}_{c1}|H\rangle^{L}_{c2}+|V\rangle^{S}_{c1}|V\rangle^{S}_{c2}).
\end{eqnarray}
Here the PC$_{A}$ is the Pockels cell \cite{pc}. It precedes the polarization
interferometers and coordinates by the time reference. It makes the PC$_{A}$ work
only at a time corresponding to the scheduled arrival of the early
state of the photons, performing the transformation
$|V^{S}\rangle\leftrightarrow|H^{S}\rangle$.

Finally, after the second Mach-Zehnder (M-Z) interferometer, the whole
state becomes
\begin{eqnarray}
&&\frac{1}{\sqrt{2}}(|H\rangle^{L}_{c1}|H\rangle^{L}_{c2}+|V\rangle^{S}_{c1}|V\rangle^{S}_{c2})\nonumber\\
&\rightarrow&\frac{1}{\sqrt{2}}(|H\rangle^{LS}_{c1}|H\rangle^{LS}_{c2}+|V\rangle^{SL}_{c1}|V\rangle^{SL}_{c2})\nonumber\\
&=&\frac{1}{\sqrt{2}}(|H\rangle_{D1}|H\rangle_{D2}+|V\rangle_{D1}|V\rangle_{D2}).
\end{eqnarray}

It is  shown that they will obtain the maximally entangled state $|\Phi^{+}\rangle$ by detecting the photons with the single-photon detectors
D$1$ and D$2$. The item $|V\rangle_{a1}|V\rangle_{a2}\otimes\frac{1}{\sqrt{2}}(|S\rangle_{a1}|S\rangle_{a2}+|L\rangle_{a1}|L\rangle_{a2})$ will be
reelected by two PBSs and become $|H\rangle_{d1}|H\rangle_{d2}\otimes\frac{1}{\sqrt{2}}(|S\rangle_{d1}|S\rangle_{d2}+|L\rangle_{d1}|L\rangle_{d2})$. It can also
become the maximally entangled state $|\Phi^{+}\rangle$ and finally be detected by D3 and D4.

The other states  $|\Phi^{-}\rangle|\Phi\rangle_{T}$,  $|\Psi^{\pm}\rangle|\Phi\rangle_{T}$ can  evolve in the same way. For example, in the first item of  $|\Phi^{-}\rangle|\Phi\rangle_{T}$, after
the decoder setup, the item
$|H\rangle_{a1}|V\rangle_{a2}\otimes\frac{1}{\sqrt{2}}(|S\rangle_{a1}|S\rangle_{a2}+|L\rangle_{a1}|L\rangle_{a2})$
becomes
\begin{eqnarray}
&&|H\rangle_{a1}|V\rangle_{a2}\otimes\frac{1}{\sqrt{2}}(|S\rangle_{a1}|S\rangle_{a2}+|L\rangle_{a1}|L\rangle_{a2})\nonumber\\
&\rightarrow&|H\rangle_{c1}|V\rangle_{d2}\otimes\frac{1}{\sqrt{2}}(|S\rangle_{c1}|S\rangle_{d2}+|L\rangle_{c1}|L\rangle_{d2})\nonumber\\
&\rightarrow&|H\rangle_{c1}|H\rangle_{d2}\otimes\frac{1}{\sqrt{2}}(|S\rangle_{c1}|S\rangle_{d2}+|L\rangle_{c1}|L\rangle_{d2})\nonumber\\
&\rightarrow&\frac{1}{\sqrt{2}}(|V\rangle_{c1}|V\rangle_{d2}\otimes|S\rangle_{c1}|S\rangle_{d2}\nonumber\\
&+&|H\rangle_{c1}|H\rangle_{d2}\otimes|L\rangle_{c1}|L\rangle_{d2})\nonumber\\
&=&\frac{1}{\sqrt{2}}(|H\rangle^{L}_{c1}|H\rangle^{L}_{c2}+|V\rangle^{S}_{c1}|V\rangle^{S}_{c2})\nonumber\\
&\rightarrow&\frac{1}{\sqrt{2}}(|H\rangle^{LS}_{c1}|H\rangle^{LS}_{d2}+|V\rangle^{SL}_{c1}|V\rangle^{SL}_{d2})\nonumber\\
&=&\frac{1}{\sqrt{2}}(|H\rangle_{D1}|H\rangle_{D4}+|V\rangle_{D1}|V\rangle_{D4}).
\end{eqnarray}
 The item
$|V\rangle_{a1}|H\rangle_{a2}\otimes\frac{1}{\sqrt{2}}(|S\rangle_{a1}|S\rangle_{a2}+|L\rangle_{a1}|L\rangle_{a2})$
can also be used to obtain the
maximally entangled state $|\Phi^{+}\rangle$ in the output modes
D2D3, respectively. Here  we only
discuss the case that the state is in the mode of $a1a2$ with the
probability of $\frac{1}{4}$. Other cases can be discussed in the
same way. We should point out that, the state in the modes $a1b2$
and $a2b1$ can be purified to
$\frac{1}{\sqrt{2}}(|HH\rangle-|VV\rangle)$.

\section{Discussion}
 So far, we have
discussed the whole purification process. It is straightforward to extend this ECP to the case of multipartite state  $\frac{1}{\sqrt{2}}(|HH\cdots H\rangle+|VV\cdots
V\rangle)$. They first generate the state $|HH\cdots H\rangle\otimes\frac{1}{\sqrt{2}}(|SS\cdots S\rangle+|LL\cdots L\rangle)$. Then all the parties
use the same setup shown in Fig. 2 to purify the polarization part. Finally they will obtain the maximally polarization entangled state with the same success probability of 100\%.
 From above explanation, the realization of
this EPP is essentially based on the robust time-bin entanglement. Several experiments described above
showed that it is indeed robust for transmission in optical fiber. Actually,  photons have the same
transmission speed in the optical channel. The state $\frac{1}{\sqrt{2}}(|S\rangle|L\rangle+|L\rangle|S\rangle)$ comes from a bit-flip error
 never occur. On the
other hand, the phase-flip error also does not exist. In the
practical transmission, the phase-flip error mainly comes from the
fluctuation of the path length. For example, in the DLCZ protocol
for quantum repeaters, the single-photon entanglement is in the form
of
$\frac{1}{\sqrt{2}}(|01\rangle+e^{i\theta_{AB}}|10\rangle)$\cite{DLCZ}.
The phase $\theta_{AB}=\theta_{A}-\theta_{B}$ denotes the difference
of the phase shifts in the left and the right sides of
channel\cite{Simon1,sangouard}. Here, path length fluctuations do not  lead
to the phase flip. Suppose a phase fluctuation occurs, it will make
the two photons in the $|SS\rangle$ time-bin become
$|S\rangle\otimes e^{i\theta}|S\rangle$, the same effect will be on
the later time-bin state, which make $|LL\rangle\rightarrow
|L\rangle\otimes e^{i\theta}|L\rangle$. The  phase $\theta$ will
become the global phase for the whole state and can be omitted. Thus,
the state such as $\frac{1}{\sqrt{2}}(|SS\rangle-|LL\rangle)$ does
not appear.

Local operations and classical communication cannot
increase entanglement. The entanglement purification can be regarded
as the transformation of the entanglement. In the previous protocol, they need two pairs of less-entangled
state in each purification process, which means that the entanglement is transformed from the
target pair to the source pair. It can be considered as the
transformation between the same kind of degree of freedom.
Surprisingly, Simon and Pan first pointed out that  the entanglement purification can be happened in
different degree of freedom \cite{Simon}. Using the entanglement encoded in other degrees of freedom provides
us a good way to perform the purification. Interestingly, one of the  significant features of these kinds of EPPs is that it does not require
the initial polarization part of the state to be entangled. It makes these kinds of EPP more like the protocol of entanglement
distribution \cite{distribution}.

 It is interesting to
compare the present protocol with the performances of the previous
EPPs. The present protocol has several advantages. First,  only one pair of state is required, while in the conventional  EPPs,
they always need two pair of low quality states.  Second, we can get a maximally
entangled pair, and the other conventional EPPs only improve the quality of
the mixed state. They should consume a large number of low quality mixed states to obtain a small number of high quality mixed states.
Compared with the other deterministic EPPs, this protocol does not resort to the hyperentanglement. Moreover, the time-bin entanglement
is more robust than the entanglement encoded in the other degrees of freedom, which makes this protocol extremely suitable in practical application.

\section{Conclusion}
In conclusion, we have presented a determinate purification protocol
for polarization entanglement state.
The success probability for this protocol is 100\%, in principle.  We do not need
large pairs of less-entangled states. As the time-bin entanglement is robust than the entanglement encoded in other degrees of freedom,
this EPP may have its practical application in current long-distance quantum communication.

\section{ACKNOWLEDGMENTS}
This work is supported by the National Natural Science Foundation of China (Grant No. 11104159), the Natural Science Research Project of Universities of Jiangsu
Province, China (Grant No. 13KJB140010), and the Priority Academic Development Program of Jiangsu Higher Education Institutions, China.

\end{document}